# A Dynamic Tracing Model for Agile Software Product Lines Domain Engineering from Features to Structural Elements: An Approach Based on Dynamic Routing

Zineb Mcharfi[1], Bouchra El Asri[2], and Abdelaziz Kriouile[3]

[1] SIME Laboratory, Mohammed V University in Rabat
Rabat, Morocco
zineb.mcharfi@gmail.com

[2] SIME Laboratory, Mohammed V University in Rabat
Rabat, Morocco
elasri@ensias.ma

[3] SIME Laboratory, Mohammed V University in Rabat
Rabat, Morocco
kriouile@ensias.ma

**Abstract**
Even if the benefit of implementing Software Product Lines is well established, adopting such a large system is still a complex choice to make: it is hard to implement, needs a good knowledge of market growth and a clear vision of the enterprise objectives for long term. Therefore, many companies remain unwilling to adopt such an approach, unless they gain flexibility and get guarantees. Flexibility can be achieved by adopting an Agile Software Product Line approach, to make sure changes are rapidly implemented and product adapted to market evolution. Guarantees can be acquired by tracing elements and the relations between them. However, tracing in Agile Software Product Line context still needs to gain maturity as it is costly and therefore rarely adopted. In this paper, we discuss the added value of traceability for Agile Software Product Lines, and present our tracing model inspired from dynamic network routing.

## 1. Introduction

Considering market growth and competitiveness, companies try to achieve mass customization with lower costs, reduce time to market, and insure product quality while getting customer satisfaction. From a software engineering point of view, Software Product Lines (SPL) is a promising concept that helps dealing with those challenges [1][2].

However, in some business environments, SPL may not be reactive enough compared to market growth. In fact, designing an SPL requires deploying considerable efforts and time in order to speculate on future products and functionalities that may be needed. Also, the Return On Investment (ROI) of those efforts might be very small in a volatile market [3]. Those constraints pushed developers and researchers to try improving SPL in order to gain flexibility, which led to the concept of Agile Product Line Engineering (APLE) [4][5][3][6].

Many researchers worked on the feasibility of combining SPL and Agile Software Development (ASD) [4][5][3][6], as both of them share the same objective of increasing productivity and software quality while optimizing production time, even if they present differences in the concept and practices [4]. Traceability might be considered as one of the challenging points when combining SPL and agility; the former, because of its complexity and need to manage variability, requires traceability documentation to assure consistency of the links between artifacts and facilitate changes implementation [2], while the latter advocates less use of documents [7].

In a previous study [26], we demonstrated the benefits of adopting a targeted tracing approach for SPL ROI. The present work is a continuity of the previous one, as we propose a targeted tracing model for ASPL. Our approach is based on similarity we established between ASPL architecture and behaviour, and dynamic routing protocols in IP networks.

The remainder of this paper is structured as follow: in Sections II, we introduce SPLs and their combination with agility. In section III we discuss traceability challenges in SPLs, agile project and ASPLs. We present our contribution in Section IV and illustrate it with a case study in Section V, before concluding in Section VI.

## 2. Background and motivations

In this section, we will first briefly introduce SPL and ASD, before presenting ASPL.

### 2.1 A quick look at Software Product Lines

SPL is "a set of software-intensive systems that share a common, managed feature set satisfying a particular market segment's specific needs or mission and that are developed from a common set of core assets in a prescribed way" [1]. It is used by organizations that produce numerous products answering specific needs but having many components in common. Those common components (e.g., architecture, requirements, test plans, schedules, budgets and processes description) are called "core assets". Adopting a SPL approach allows producing new systems by reusing the existing ones, in an organized manner.

SPL is a combination of three major interacting elements [1][8]: (1) core asset development or Domain Engineering (DE), (2) product development or Activities Engineering (AE) and (3) technical and organizational management that orchestrates those two activities.

SPL is by far considered as an up-front, proactive (in opposite to reactive) reuse demarche [9]: it is based on a production plan, involves both technical and organizational management, is a direct result of the organization strategy, and it is used to reach predictable results.

### 2.2 Agility and Software Product Lines: A big challenge

As explained earlier, SPLE is based on up-front designing with heavy processes and significant efforts. It helps answering planned changes. However, in unstable environments with rapidly changing conditions, the investment in SPL might be pricey [3]. On the other hand, ASD seeks to satisfy customer requirements in a reactive way, promoting continuous discussion with the customer, and avoiding up-front developments.

According to Díaz et al. [3] and Ghaman et al. [5], the combination of SPL and ASD principles allows eliminating long term investment in up-front design, especially in volatile markets where it would represent a non-profitable investment in the long term with huge losses due to no-longer useful core assets or never-used ones. It allows also dealing with situations where there is lack of knowledge about domain engineering or where no speculation can be made.

Agility has been combined to SPL at different stages: (i) In planning and scoping by using a collaborative approach [10] or by implementing an agile scoping process [11] with iterative and incremental phases, continuous communication with customers and the use of user stories; (ii) in architecting, as in [12], by implementing a process to assist architects while building the PL in an agile demarche; (iii) in product derivation as demonstrated in [13] and [14] in small environments through an Agile process model; (iv) in variability management as in [15] [16] who used a Test Driven Development (TDD) method in an Extreme Programming environment to deal with agility in SPL; (v) in Product Line evolution as in [17] which described the application of agility for SPL evolution through a case study, using Composing Feature Models (CFM).

## 3. Traceability challenges in Agile Software Product Lines

In such a complex environment (i.e., ASPL), where we have to manage variability in a constantly evolving context, it is very important to insure traceability along the software development process.

However, based on literature analysis, we notice that very few researches deal explicitly with the problematic of traceability in ASPL, even though managing traceability is very important in such evolving environments. Therefore, we chose to discuss traceability in ASPL, knowing the challenges that it presents.

### 3.1 Traceability in Software Product Lines

Traceability helps follow the components' life, links the different software artifacts, from requirements to source codes and backwards and, in a larger scale, helps verify that all requirements have been implemented and the artifacts documented [18]. It is also a way to consider different architectural choices and identify errors, and facilitate communication between stakeholders [19]. Traceability is very helpful when it comes to maintenance and evolution as it allows analysing and controlling the impact of changes [20]. However, SPLs add complexity to traceability due to their reuse characteristics and variability management [21].

From a conceptual view, Berg et al. [22] proposes a variability model to deal with traceability in SPL and consider that, in addition to the two dimensions of traceability in simple software (i.e., phases of development and levels of abstraction), there is a need to add variability as a third dimension in SPL. Anquetil et al. [14][16] added a fourth dimension, namely evolution, to link between the different versions of every artifact, and a fifth one, versioning, to trace the components' changes in time.

Traces can also be seen as links that define a path between SPL elements. Three different traceability issues are defined in [23]: (1) Between features and structural elements, at both DE and AE levels, to allow identifying components needed to deliver desirable feature in DE and to localize the impact of a change on product components in AE; (2) Between DE and AE levels, to keep traces between the product line and how it is implemented in the different derived products (components used, variants chosen,...); (3) Between concrete and generic solutions: from an architectural point of view, to trace the relationships between patterns (example architectural patterns) and the concrete solutions.

3.2 Traceability in Agile Projects

At first sight, traceability and agility might not be compatible as the former generates additional effort and documentation, while the latter is known to be a lightly documented method.

However, some agile projects, especially those related to critical domains or complex large-scale projects, need to give credibility to their resulting products, in which case tracing is needed [24].

In order to create and maintain traces while remaining agile, tracing has to be light and quickly adapts to the environment changes [24]. Three main tracing approaches in agile environment are found in literature [24]: (i) Traceability Information Model (TIM) [30], that is "a graph defining the permissible trace artifact types, the permissible trace link types and the permissible trace relationships on a project, in order to address the anticipated traceability- related queries and traceability-enabled activities and tasks". It helps link user stories to acceptance tests, and concerns a superficial traceability granularity, where code is considered as a whole, without defining traces between its different classes [24]. This approach can be used for small-size projects and those where there is no necessity for explicit links to code classes [24]. (ii) Just-in-time Traceability (JITT), which is an automated traceability method generally used in large projects where developers have no sufficient knowledge of the code and relations between features [24]. Traces are created on demand, when needed. This approach can be helpful when developers are not familiar with the project, especially in agile environment where requirements change continuously. (iii) Lean Traceability, which consists on lightening tracing constraints to match agility aims, and one lean tracing possibility is tracing when needed, at the granularity level intended [29]

3.3 Traceability in Agile Software Product Lines

As mentioned earlier, only few works on ASPL deal with tracing issues in their solutions. In [15] [16], Ghanam works on variability management in an XP SPL environment. He used produced Acceptance Tests (AT) to provide traceability information. The work of [25] illustrates a combination of workflow and Web Services to create a web application. The PL architecture is based on interrelated components and their instantiation is made agile by the use of a graphical interface. Traceability is therefore assured by the use of workflow and the graphical interface.

# 4. Outlook and contribution

In earlier works [26] [27], we discussed ROI of adopting a specific tracing strategy (full, adhoc or targeted) at different SPL levels (DE, AE and maintenance). According to this study, traceability is costly at DE, while implementing traces. However, the costs can be reduced by adopting a targeted traceability. Therefore, we aim in our study to establish a model that helps optimizing traceability in DE, and we'll focus on links between features and structural elements (architectural model).

4.1. Basic concepts

- Network routing protocols

According to Cisco definition [28], "routing is the act of moving information across an internetwork from a source to a destination". Thus, determining the path between source and destination is an important routing element, besides network components.

Many routing protocols exist and they use metrics (example: next hop) to determine the best path to take. They are also based on algorithms that use routing tables to store information about adjacent components and routing paths. A routing algorithm can be either static or dynamic. Static routing algorithms are based on fixed tables filled by network administrator. Administrator intervention is needed to transcribe any network change. Static routing tables rarely change and are used in simple networks, with predictable traffic [31]. For networks with several routers and where changes occur frequently, dynamic routing is needed [31].

Based on routing protocol, routing table is updated when network topology changes, and the new situation is broadcasted through the network so as the other routers are informed to update their own routing tables.

A dynamic routing protocol is composed of [32]:
- Data structures: like tables or databases to store information concerning routing operations
- Routing protocol messages: to exchange information with network elements like neighbours
- Routing algorithm: used for best path determination

Dynamic routing protocols can be classified into two big families: Distance vector and link-state routing protocols. In distance vector protocol, each router knows its neighbours (routers) and networks it can reach throughout those neighbours [33]. The router sends periodically a broadcast message to its neighbours to detect network updates.

In link state routing are based on Dijkstra's shortest path first (SPF) algorithm [34] which consist on "finding the shortest paths between nodes in a graph" [35] The SPF algorithm helps generating a map of network topology that is used by routers to reach the desired destination node.

4.2. Our contribution: Dynamic tracing for ASPL

As mentioned earlier, we are interested in establishing a tracing methodology to promote and optimize traceability in ASPL, especially at DE phase, between features and architectural model.

In order to reach this goal, we based our work on network dynamic routing. In fact, tracing can be assimilated to drawing a path (i.e. routing) between two points (i.e. network elements). Connections between features and components define a network.

In SPL in general and more particularly in ASPL, links between features and architectural components are many to many relations: a feature can be implemented by one or more component and a component can instantiate one or more features (Fig. 1).

The elements of our "ASPL network" are features and components. We divide that network into subnetworks, in order to have 1 to 1 relations between features set and subnetworks set. A component may belong to one or more subnetworks. Similarly to Internet, each subnetwork from our "ASPL network" is identified by a specific address. Components that belong to more than one subnetwork represent the junctions between subnetworks and therefore will be assimilated to routers in Internet networks. They have 2 or more addresses (Fig. 2).

Addresses will help tracing elements. In fact, like routing Internet packages from source to destination, tracing consists in finding a path from point A to point B, which could be done using addresses and networking algorithms.

As discussed in the previous paragraph, routing in Internet networks is either static or dynamic. For the tracing solution we propose, we adopted a dynamic approach. Indeed, ASPL are complex large scale systems. Also, in such an agile environment, maintaining heavy documentation is avoided. Another argument that justifies our choice is the main characteristic of SPL: variability. In fact, one of the most important characteristics of SPL is variation points, and adopting a dynamic tracing solution will help optimize traces generation and maintenance: we assign an address only for the variation point. No address is attributed to variants at DE phase but only when instantiating, at AE step.

Also, an SPL is composed from common artifacts that are shared by all the products and other specific to some of them. With our dynamic tracing solution, traces between common artifacts can be generated in DE, while those for specific components might be created in AE and whenever they are needed.

As explained in the previous paragraph, there are many algorithms that can be adopted for a dynamic network routing. We chose to use Distance vector algorithm for our solution, as we look for reducing tracing complexity in agile and large scale environment.

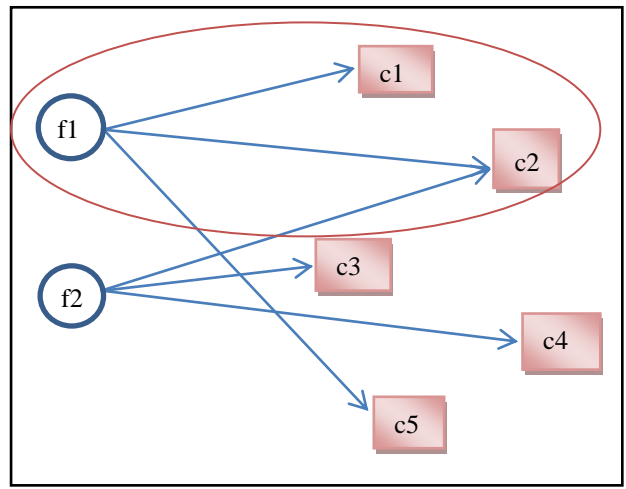

Fig. 1 Links between features and architectural components in a SPL.

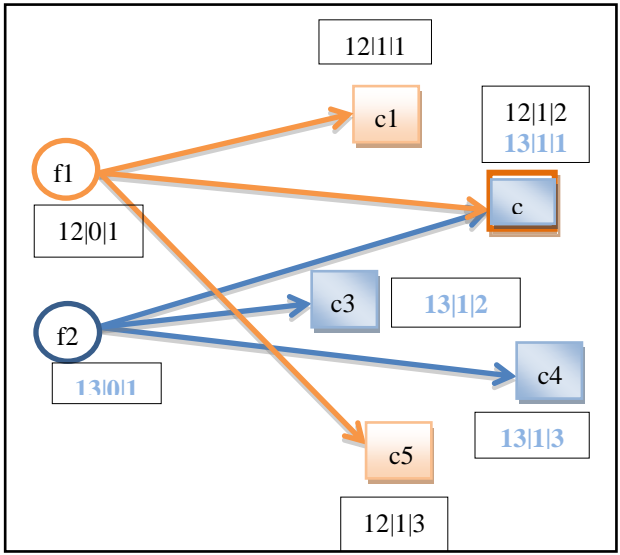

Fig. 2 Decomposition of the SPL into subnetworks and components' addresses allocation.

The proposed solution will allow:
- Tracing from feature F to component C
- Detecting architectural changes (new feature/component; elements deleted) and adjust traces dynamically
- Identifying products that contain specific elements (to study change impact).

From an agile point of view, our solution allows some flexibility and eases maintenance since we only record router address and element address for each feature and simple component, and routers and neighbours' addresses for components that serve as routers. Also links are dynamically updated; no specific maintenance routine is needed.

As for IP addresses, we setup addressing in our solution so as to encapsulate the maximum of interesting information. Each address is composed of 3 parts: subnetwork id, element category (feature = 0/component = 1), and element id (Fig. 2).

Next section will present how our tracing solution can be implemented when creating an ASPL for projects specification management.

## 5. Case study: Tracing in a platform for telecommunication operator offers, from conception to launch

In telecommunication market, offers evolve continuously to meet the customers' needs. However, the process from offer conception to offer launch rarely changes, and despite their content differences, telecom offers present similarities and proper instantiations. This led us to choose building an SPL for telecom offers creation and maintenance.

Also, as telecom market is rapidly growing, offers are continuously changing. Therefore, a rigid creation solution might be costly in time and resources. Adopting an ASPL approach can be considered as an optimal solution.

In the same context of rapidly evolving telecommunication offers, tracing is a key element as it helps identifying specific changes in an offer, impacted elements and even similar offers that might be impacted. This early identification of changes impact is precious as long as most changes are directed by the regulator and must be applied as soon as possible, before deadline.

Given the foregoing, we decided to apply our tracing model in the context of establishing a platform for telecommunication offers, from conception to launch.

Fig. 3 describes, through feature and component models, the process by which a telecommunication operator designs, setups and launches a new offer.

Starting from this figure and by applying the approach we proposed above, we identify 3 subnetworks. A subnetwork related to the feature "Hierarchy validation", linked to the components "Network parameters" and "offer validation"; a second subnetwork for "Telecom regulator validation" feature, where the latter uses the components "offer validation" and "tariff", and a last feature "offer elaboration" that we'll take a look at.

In fact, many components are needed to implement the "offer elaboration" feature: "Customer category", "flow", "tariff", "offer duration" and "Network parameters". The latter component is shared by many features, thus, it will be the routing element in our case study (Fig. 3). While developing the platform, two addresses are assigned to "Network parameters" component: 12|1|2 and 13|1|1: It is a component (element category = 1), part of subnetworks 12 and 13, and has, respectively, the ids 4 and 2.

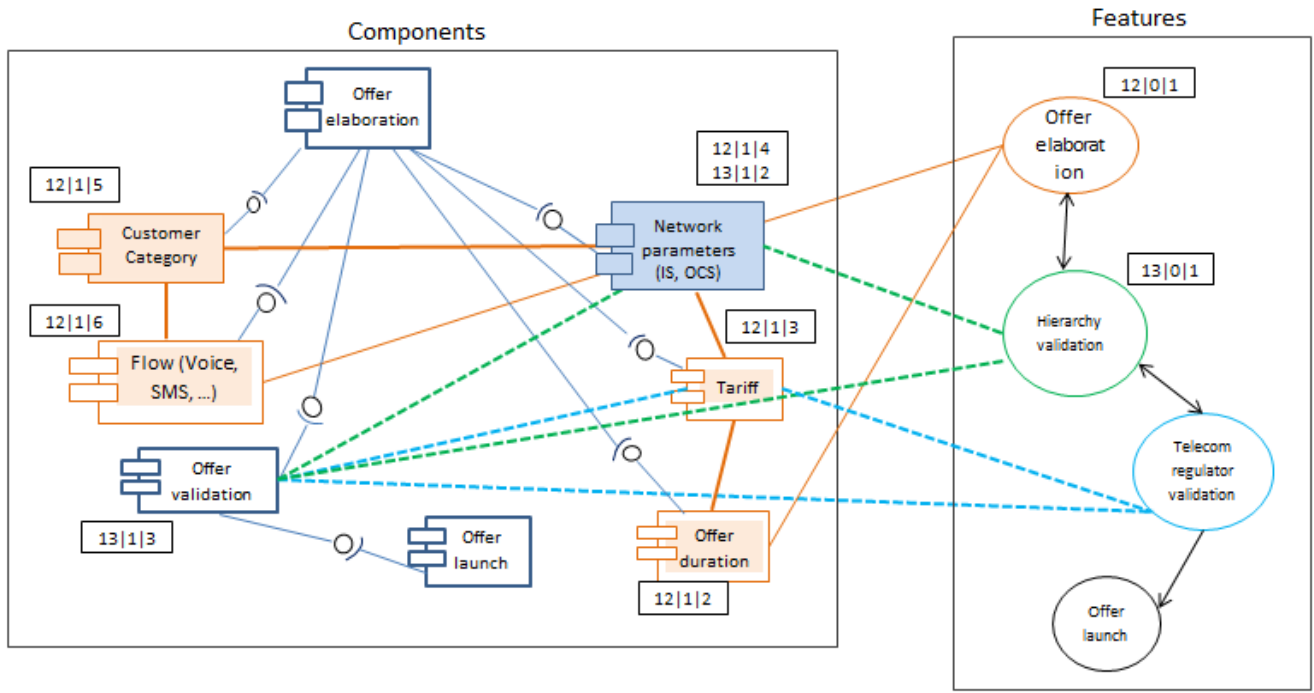

Fig. 3 Process by which a telecommunication operator designs, setups and launches a new offer.

One possible change in the context of telecommunications offers is hierarchy validation. In fact, the hierarchy might, following the request of regulator, modify its offer validation and ask for new conditions, especially in tariff and duration.

Starting from the feature "Hierarchy validation", we look for the rooting element in the same subnetwork (subnetwork id = 13), "Network parameters". As it is our network router, and according to our routing approach, it knows its neighbours, therefore, it will recognize the link to the element "tariff" (1$^{st}$ level neighbour) in subnetwork 12. It also recognizes the "duration" component in the same subnetwork.

Another possible use of tracing in this context might be for architecture change, especially when deleting a component. In our case study, one possible change is to propose, at the request of hierarchy, offers regardless of customer category. In that case, once the component "Customer Category" is deleted from our SPL, the routing component "Network parameters" will detect the change and trace links will be adjusted dynamically.

## 6. Conclusion

The benefits of tracing for ASPL don't need to be demonstrated anymore. However, due to the complexity of establishing tracing approaches in such complex and large context, traceability is rarely given importance during the design and the development of the ASPL.

Therefore, we proposed a dynamic tracing algorithm to deal with the issues of traceability in ASPL environments. Our algorithm is inspired from dynamic routing algorithms because we noted that it shares similarities with tracing for ASPL.

Through our study, we tried to suggest an approach that deals with traceability while taking into consideration development time and costs. Therefore, the next step in our work is to evaluate the cost of the proposed dynamic tracing compared to other ASPL tracing approaches.

## References

[1] L. M. Northrop, "SEI's software product line tenets," IEEE Softw., vol. 19, no. 4, pp. 32–40, 2002.
[2] K. Pohl, G. Böckle, and F. Van Der Linden, Software product line engineering. 2005.
[3] J. Díaz, J. Pérez, P. P. Alarcón, and J. Garbajosa, "Agile Product Line Engineering - A Systematic Literature Review," Softw. Pract. Exp., vol. 41, no. 8, pp. 921–941, 2011.
[4] K. Tian and K. Cooper, "Agile and software product line methods: are they so different," 1st Int. Work. Agil. Prod. Line Eng. (APLE), collocated with 10th Int. Softw. Prod. Line Conf., 2006.
[5] Y. Ghanam, F. Maurer, and K. Cooper, "A Report on the XP Workshop on Agile Product Line Engineering," ACM SIGSOFT Softw. Eng. Notes, vol. 34, no. 5, pp. 25–27, 2009.

[6] R. Carbon, M. Lindvall, D. Muthig, and P. Costa, "Integrating Product Line Engineering and Agile Methods: Flexible Design Up-Front vs. Incremental Design," in 1st International Workshop on Agile Product Line Engineering APLE06, 2006, pp. 1–8.
[7] M. Fowler and J. Highsmith, "The agile manifesto," Softw. Dev., vol. 9, pp. 28–35, 2001.
[8] L. Northrop and P. Clements, "Software Product Lines," Carnegie Eng. Inst., pp. 1–105, 2005.
[9] C. Krueger, "Eliminating the adoption barrier," IEEE Softw., vol. 19, no. 4, pp. 29–31, Jul. 2002.
[10] M. a. Noor, R. Rabiser, and P. Grünbacher, "Agile product line planning: A collaborative approach and a case study," J. Syst. Softw., vol. 81, no. 6, pp. 868–882, Jun. 2008.
[11] M. Balbino, E. S. De Almeida, and S. Meira, "An Agile Scoping Process for Software Product Lines," in SEKE 2011 - Proceedings of the 23rd International Conference on Software Engineering and Knowledge Engineering, 2011, pp. 717–722.
[12] J. Díaz, J. Pérez, and J. Garbajosa, "Agile product-line architecting in practice: A case study in smart grids," Inf. Softw. Technol., vol. 56, no. 7, pp. 727–748, Feb. 2014.
[13] P. O'Leary, M. A. Babar, S. Thiel, and I. Richardson, "Product Derivation Process and Agile Approaches: Exploring the Integration Potential," in Proceedings of 2nd IFIP Central and East European Conference on Software Engineering Techniques, 2007, pp. 166–171.
[14] P. O'Leary and F. McCaffery, "An agile process model for product derivation in software product line engineering," J. Softw. Evol. Process, 2012.
[15] Y. Ghanam and F. Maurer, "Extreme Product Line Engineering: Managing Variability and Traceability via Executable Specifications," in 2009 Agile Conference, 2009, pp. 41–48.
[16] Y. Ghanam, "An Agile Framework for Variability Management in Software Product Line Engineering," 2012.
[17] S. Urli, M. Blay-Fornarino, P. Collet, and S. Mosser, "Using composite feature models to support agile software product line evolution," in Proceedings of the 6th International Workshop on Models and Evolution - ME '12, 2012, pp. 21–26.
[18] J. Cleland-Huang, O. Gotel, J. H. Hayes, P. Mäder, and A. Zisman, "Software Traceability: Trends and Future Directions," in 36th International Conference on Software Engineering (ICSE), 2014.
[19] N. Anquetil, U. Kulesza, R. Mitschke, A. Moreira, J.-C. Royer, A. Rummler, and A. Sousa, "A model-driven traceability framework for software product lines," Softw. Syst. Model., vol. 9, pp. 427–451, 2010.
[20] Y. C. Cavalcanti, I. do Carmo Machado, P. A. da Mota, S. Neto, L. L. Lobato, E. S. de Almeida, and S. R. de Lemos Meira, "Towards metamodel support for variability and traceability in software product lines," in Proceedings of the 5th Workshop on Variability Modeling of Software-Intensive Systems - VaMoS '11, 2011, pp. 49–57.
[21] N. Anquetil, B. Grammel, I. Galvão, J. Noppen, S. Khan, H. Arboleda, A. Rashid, and A. Garcia, "Traceability for Model Driven, Software Product Line Engineering," ECMDA Traceability Work. Proc., 2008.
[22] K. Berg, J. Bishop, and D. Muthig, "Tracing Software Product Line Variability - From Problem to Solution Space," Proc. 2005 Annu. Res. Conf. South Africain Inst. Comput. Sci. Inf. Technol. IT Res. Dev. Ctries., pp. 182–191, 2005.
[23] P. Lago, H. Muccini, and H. van Vliet, "A scoped approach to traceability management," J. Syst. Softw., vol. 82, no. 1, pp. 168–182, 2009.
[24] J. Cleland-Huang, "Traceability in Agile Projects," in Software and Systems Traceability, vol. 9781447122, London: Springer London, 2012, pp. 265–275.
[25] M. Karam, S. Dascalu, and H. Safa, "A product-line architecture for web service-based visual composition of web applications," J. Syst. Softw., vol. 81, no. 6, pp. 855–867, 2008.
[26] Z. Mcharfi, B. El Asri, I. Dehmouch, A. Baya, and A. Kriouile, "Return on Investment of Software Product Line Traceability in the Short, Mid and Long Term," in Proceedings of the 17th International Conference on Enterprise Information Systems, 2015, pp. 463–468.
[27] Z. McHarfi, B. El Asri, I. Dehmouch, and A. Kriouile, "Measuring the impact of traceability on the cost of Software Product Lines using COPLIMO," Proc. 2015 Int. Conf. Electr. Inf. Technol. ICEIT 2015, pp. 192–197, 2015.
[28] "Routing Basics - DocWiki." [Online]. Available: http://docwiki.cisco.com/wiki/Routing_Basics..
[29] B. Appleton, S. Berczuk, and R. Cowham, Lean-agile traceability: Strategies and solutions, 2007.
[30] O. Gotel, J. Cleland-Huang, J. Hayes, et al. Traceability fundamentals. In: Software and Systems Traceability. Springer London, 2012. p. 3-22.
[31] Cisco press, "Dynamic versus Static Routing", http://www.ciscopress.com/articles/article.asp?p=2180210&seqNum=5.
[32] Cisco press, "Dynamic Routing Protocols", http://www.ciscopress.com/articles/article.asp?p=2180210&seqNum=4.
[33] Cisco press, "Distance Vector Dynamic Routing", http://www.ciscopress.com/articles/article.asp?p=2180210&seqNum=8.
[34] Cisco press, "Link-State Dynamic Routing", http://www.ciscopress.com/articles/article.asp?p=2180210&seqNum=11.
[35] https://en.wikipedia.org/wiki/Dijkstra%27s_algorithm

**Zineb Mcharfi** received a degree in software engineering from National High School of Computer Science and Systems Analysis (ENSIAS) in 2008. She is currently a PhD student in the IMS (Models and Systems Engineering) Team of SIME Laboratory at ENSIAS. Her research interests include Software Product Line Engineering, Agile Software Development and software traceability.

**Bouchra El Asri** is a Professor in the Software Engineering Department and a member of the IMS (Models and Systems Engineering) Team of SIME Laboratory at National High School of Computer Science and Systems Analysis (ENSIAS), Rabat. Her research interests include Service-Oriented Computing, Model-Driven Engineering, Cloud Computing, Component-Based Systems and Software Product Line Engineering.

**Abdelaziz Kriouile** is a full Professor in the Software engineering Department and a member of SI2M Laboratory at National Higher School for Computer Science and Systems Analysis (ENSIAS), Rabat. He is also a Head of the SI3M Formation and Research Unit. His research interests include integration of viewpoints in Object-Oriented Analysis/Design, Service-Oriented Computing and speech recognition by Markov models. He has directed several Ph.D. thesis in the context of Franco-Moroccan collaborations.